\documentstyle[12pt,epsfig]{article}

\newlength{\dinwidth}
\newlength{\dinmargin}
\def\msb{\overline{\rm MS}}

\def\lapproxeq{\lower .7ex\hbox{$\;\stackrel{\textstyle
<}{\sim}\;$}}
\def\gapproxeq{\lower .7ex\hbox{$\;\stackrel{\textstyle
>}{\sim}\;$}}

\def\be{\begin{equation}}
\def\ee{\end{equation}}
\def\bea{\begin{eqnarray}}
\def\eea{\end{eqnarray}}
\def\lapproxeq{\lower .7ex\hbox{$\;\stackrel{\rm <}{\sim}\;$}} 
                                                         
\def\gapproxeq{\lower .7ex\hbox{$\;\stackrel{\rm >}{\sim}\;$}} 

\def\s23{i}
\def\xx0{\frac{x}{x_0}}
\def\zx0{\frac{z}{x_0}}
\def\hs{\tilde s}

\begin{document}
\titlepage
\begin{flushright}
RAL--TR--2000--048\\
Cavendish-HEP-2000/10\\
October 2000 \\
\end{flushright}
\begin{center}
\vspace*{2cm}
{\large \bf A Variable Flavour Number Scheme for \\[2mm]
Charged Current Heavy Flavour Structure Functions}\\
\vspace*{1cm}
R.\ S.\ Thorne\footnote{Royal Society 
University Research Fellow} \\

{\it Cavendish Laboratory, University of Cambridge,
Madingley Road, Cambridge, CB3 0HE}

and

R.\ G.\ Roberts \\

{\it Rutherford Appleton Laboratory, Chilton,
Didcot, Oxon, OX11 0QX}\\

\end{center}

\vspace*{1.5cm}
\begin{abstract}

The Thorne-Roberts variable flavour number scheme (VFNS) for heavy quarks is 
presented in detail for the specific case of charged current DIS. As in 
neutral current DIS this provides a smooth extrapolation from the
fixed flavour number scheme (FFNS) appropriate at low $Q^2$ to the zero-mass
variable flavour number scheme (ZM-VFNS) appropriate as $Q^2 \to \infty$,
and differs from alternative versions of a VFNS by the definition of the 
coefficient functions at each order, and the strict ordering of the expansion 
in $\alpha_S$. However, there are subtle differences from the neutral current
case which are addressed here. We discuss both the LO and NLO expressions, 
the latter
unfortunately requiring some (minimal) modelling due to the current lack of 
some necessary ${\cal O}(\alpha^2_S)$ FFNS coefficient functions.    
 
\end{abstract}

\newpage

\noindent{\large \bf 1 {\hskip 0.5cm} Introduction.}

\medskip

In the past few years direct measurements of charm production at HERA 
\cite{h1charm,zeuscharm},
as well as the fact that the charm
structure function $F_2^c$ can be 20$\%$ or more of the total $F_2$,
have made a consistent theoretical framework for 
heavy flavour production in neutral current deep inelastic scattering 
(DIS) essential. 
For $Q^2\lapproxeq m_c^2$, where $m_c$ is the charm quark mass, 
the conventional description in terms of order-by-order in $\alpha_s$ 
coefficient functions for the production of charm in the final state
is perfectly satisfactory, but for $Q^2 >> m_c^2$ this description becomes 
potentially unreliable
due to the presence of logarithms in $\ln(Q^2/m_c^2)$
at all orders in $\alpha_s$ which ideally should
be resummed.  By changing to the alternative description where the
charm quark is treated as a parton this resummation
is automatically performed and, at the same time, a complete set of parton
densities needed to calculate other processes involving nucleons is 
obtained. It is relatively straightforward to do this by treating charm 
as a massless parton, thus obtaining the correct high $Q^2$ limit, but
more challenging to obtain a treatment which successfully includes the charm 
mass effects for $Q^2$ not too far above $m_c^2$. 

This problem was first addressed in \cite{acot}, and the term variable 
flavour number scheme (VFNS) coined for a general order-by-order 
prescription for the calculation of 
$F_2^c$ which extrapolates  from $Q^2\leq m_c^2$, to the asymptotic limit 
$Q^2/m_c^2 \to \infty$. However, while the prescription in \cite{acot} is 
certainly formally correct (an all orders proof being presented in 
\cite{collins})\footnote{Strictly speaking this scheme, as well as ours, 
applies to the total structure function $F_2$ - the explicit separation of a 
charm component $F_2^c$ becoming ambiguous beyond NLO, as discussed in 
\cite{smith}.} it is arguable that it is not the most efficient and elegant 
definition of a VFNS. In particular, the correct threshold behaviour for 
charm pair production is not precisely maintained order by order, leading to 
a lack of smoothness when one begins charm parton evolution, particularly
in $F_L^c$.
In \cite{TR} we developed a VFNS which while to all orders is 
identical to that in \cite{acot}\footnote{This is only exactly true if
there is no intrinsic charm. If there is intrinsic charm the schemes differ 
by ${\cal O}(\Lambda^2_{\rm QCD}/Q^2)$ \cite{durham}, i.e. of the order of 
the error in perturbative QCD.}, differs at any fixed order in perturbation
theory, both due to the definition of the coefficient functions and due 
to the way in which we define what a given order actually means. 
Essentially, we use the inherent freedom in the definition of the 
coefficient functions corresponding to charm partons to ensure smoothness 
of the structure functions across the transition point where we switch from 
a three to a four flavour scheme. This results in the ``Thorne-Roberts''
VFNS. 

Recently it has become apparent that the treatment of heavy flavours is also
very important in the context of charged current scattering. In particular,
a consistent method is needed in order to explain the data obtained by 
CCFR \cite{CCFR}, and resolves the long standing discrepancy
between this and the NMC muon data \cite{NMC} for $x\leq 0.01$ (see 
\cite{unki} for a presentation of the ``Physics-Model Independent'' 
treatment of this data, and \cite{kosty} for a discussion of the theoretical
issues involved). Although we outlined the way in which one treats 
charged currents in the TR scheme in the latter of \cite{TR}, and produced 
publicly available code at the time, closer examination has revealed that 
the issue is more subtle than we originally believed, and that 
the original code contained some errors. Hence, in this paper we will 
present the explicit form of the TR VFNS for charged currents, and accompany
the paper with a revised code. The paper contains our results at
leading order (LO) and next-to-leading order (NLO). The former of these is 
complete, while unfortunately the latter contains a certain amount of 
modelling since the NLO fixed flavour number scheme (FFNS) 
coefficient functions,
which would be necessary in the full prescription, have not yet been 
calculated. We find that a minimal degree of sophistication is needed in the 
modelling however.

\bigskip

\noindent{\large \bf 2 {\hskip 0.5cm} Definition  of the VFNS at LO.}

\medskip

The general formulation of the VFNS for charged current interactions was 
presented in the appendix of the former of \cite{TR}. Here we shall be  
more explicit, and work order by order.  
We consider the process $W^+ \rightarrow c, \bar s$ or $W^+ \rightarrow c, 
\bar d$. The whole of the variable flavour number scheme is based on the 
idea that ignoring any intrinsic heavy flavour (as we shall do in this paper)
the partons in the 4-flavour scheme are related to those in the 
3-flavour scheme
by
\be
f^4_b(z,\mu^2,m_c^2/\mu^2) = A^{ba}(\mu^2/m_c^2) \:\otimes\: 
f^3_a(\mu^2),\label{eq:partons}
\ee  
where the $A^{ba}(\mu^2/m_c^2)$ are perturbatively calculable matrix 
elements which are known to ${\cal O}(\alpha_S^2)$ \cite{buza}. By using 
the exact equivalence of the total structure function calculated in either 
the FFNS or the VFNS, i.e. 
\be
F_i(x,Q^2,m_c^2)= C^{FF}_{a} \:\otimes\: f^3_a(\mu^2) \equiv  
C^{VF}_{b} \:\otimes\: f^4_b(z,\mu^2,m_c^2/\mu^2),\label{eq:totf}
\ee
and using eq(\ref{eq:partons}) one obtains the VFNS coefficient functions 
in the implicit form  
\be 
C_b^{VF}(z,Q^2/\mu^2,m_c^2/\mu^2) = C_a^{FF}(Q^2/\mu^2,m_c^2/\mu^2)
\:\otimes\: \biggl [ A^{ba}(\mu^2/m_c^2) \biggr ]^{-1}.
\label{eq:corrcoeff}
\ee 
Since the index $b$ runs over one more value than index $a$ the above 
equation does not have unique solutions, and there is an inherent freedom in 
the definition of the VFNS coefficient functions to which the 
parton distributions are completely insensitive. In \cite{acot} they were 
calculated explicitly in $\msb$ scheme using charm quarks in the initial 
state, and all collinear divergences systematically removed. 
The same direct definition of coefficient functions is used in \cite{smith}.
However, at fixed order this definition can lead to single charm quark 
production below the real threshold 
for pair production, as well as a lack of smoothness when switching from 
3 to 4 flavours (depending on renormalization/factorization scale),
as clearly seen in fig. 18 of \cite{smith}. Hence,
we use an alternative method to define the coefficient functions while using 
the same parton distributions. We use the freedom to redefine coefficient 
functions while still satisfying eq(\ref{eq:corrcoeff}) and solve  
eq(\ref{eq:corrcoeff}) order by order, removing the freedom  by imposing the 
continuity of $dF_i(x,Q^2)/d\ln Q^2$ (in the gluon sector) across the 
transition point. This then guarantees the correct threshold behaviour 
in each coefficient function, and smoothness when switching flavour number. 
The difference between this scheme and \cite{acot} and \cite{smith} is 
effectively a change of factorization 
scheme such that coefficient functions differ by ${\cal O}(m_c^2/Q^2)$, but
where the parton distributions are 
identical in the two schemes. Hence, our VFNS uses the standard 4-flavour 
$\msb$ partons.  

The case of charged current scattering is made more complicated than that
for neutral currents by the fact that the heavy quark is often produced 
along with, or from, a light quark, rather than in a heavy 
quark-antiquark pair. Hence, within the FFNS it is not the production of 
the charm quark which vanishes at zeroth order in $\alpha_S$. In fact at 
LO the charm quark structure function is given by 
\bea
F_2^{c,LO}(x) &=& 2 \;[\;\cos^2 \theta_c \;\;\xi s(\xi) 
+ \sin^2 \theta_c \;\;\xi d(\xi) \;] \nonumber \\
xF_3^{c,LO}(x) &=& 2 \;[\;\cos^2 \theta_c \;\;x s(\xi) 
+ \sin^2 \theta_c \;\;x d(\xi) \;] 
\label{eq:n1}
\eea
where $\xi = x/x_0$, $x_0 = 1/(1+\epsilon) $ and $\epsilon = m_c^2/Q^2$.
The partons being functions of $\xi$ rather than $x$ due to the need to 
put the charm quark on mass-shell. From now on we will denote
$[\;\cos^2 \theta_c \;\;\xi s + \sin^2 \theta_c \;\;\xi d \;]$ by $\hs$.
At zeroth order it is the production of the weak eigenstate conjugate to 
$c$, i.e. $\bar{\hs}$, which has zero production cross-section in the FFNS.  
In this scheme the LO contribution to $\bar{\hs}$ production 
is (we choose $\mu^2=Q^2$)
\be
F_{\s23}^{\bar{\hs}}(x) = 2 \left ( \frac{\alpha_S}{4\pi} \right ) \int_x^{x_0}
dz\; C_{\s23 ,g}^{(1)FF}(z,\epsilon) \; \tilde g(x/z)
\label{eq:n3}
\ee
where $\tilde g(x) = x g(x)$, and in this paper we consider the cases $i=2$ and
$i=3$. The fixed flavour coefficient functions on the 
rhs of eq(\ref{eq:n3}) are related to the the $W^+ g \rightarrow c \bar s$ 
coefficients given by \cite{gkr}
\bea
C_{2,g}^{(1)FF}(z,\epsilon) &=& \frac{2}{x_0}\; H_2^g(\frac{z}{x_0}) 
\nonumber \\
C_{3,g}^{(1)FF}(z,\epsilon) &=& 2\; H_3^g(\frac{z}{x_0}), 
\label{eq:n4}
\eea
which are an update of those in \cite{gott} to account for the correct
counting of gluon helicity states in $D=4+2\epsilon$ dimensions.
The factor of $2$ is just our convention while the factors of $x_0$ come from
a change of variables in the integration defining the convolution compared to
\cite{gkr}.

Above the transition point\footnote{Throughout we ignore reference to the
number of flavours concerning $\alpha_S$. However, $\alpha_S(Q^2)$ 
does change across the transition point as discussed for the neutral current 
case in \cite{TR}.}, which as before we choose for convenience to be
$Q^2=m_c^2$, one can produce $\bar{\hs}$ quarks directly from initial state 
charm quarks, i.e. at LO the ($Q^2$-dependent) LO expression is 
\be
F_{\s23}^{\bar{\hs}}(x) = \int_x^{x_0}
dz\; C_{\s23 ,\bar{\hs} c}^{(0)VF}(z,\epsilon) \; \tilde  c(x/z).
\label{eq:n3a}
\ee
In principle we now impose the continuity of $\frac{d F_{\s23}^{\bar{\hs}}(x)}
{d\ln Q^2}$ at first order in $\alpha_S$ to obtain
\be
\frac{dC_{\s23 ,g}^{(1)FF}(z,\epsilon)}{d \ln Q^2}= 
C_{\s23 ,\bar{\hs} c}^{(0)VF}(\epsilon) \:\otimes\: P^{(0)}_{qg},
\label{eq:naive}
\ee
where $P_{qg}^{(0)}(z) = 1/2(z^2 +(1-z)^2)$, i.e. the LO quark-gluon 
splitting function. 
As shown in \cite{TR} one may invert eq(\ref{eq:naive}), more easily by 
considering the ultimate convolution with the charm density, obtaining   
\bea
C_{\s23,\bar{\hs}c}^{(0)\;VF}(\epsilon)\otimes c(Q^2)
&=& - \int_{x}^{x_0} dz \, \frac{d C_{\s23 ,g}^{(1)FF}(z,\epsilon)} 
{d \ln Q^2}
\biggl({x\over z}\biggr)^2 \: \frac{d c(x/z,Q^2)}
{d (x/z)} \nonumber \\
&+& 2\int_{x}^{x_0} dz \, \frac{d C_{\s23 ,g}^{(1)FF}(z,\epsilon)} 
{d \ln Q^2}
\:{x\over z} \: c(x/z,Q^2) \nonumber \\
&-& 2\int_{x}^{x_0} dz \, \frac{d C_{\s23 ,g}^{(1)FF}(z,\epsilon)} 
{d \ln Q^2}
\int_{{x/z}}^1 d z'\, r(z') \: {x\over zz'}
\: c(x/zz',Q^2),
\label{eq:conv} 
\eea
where $r(z)$ is given by
\be
r(z) = z^{1 \over 2}\biggl[ \cos\Bigl({\sqrt 7 \over 2}\ln 
{1\over z}\Bigr) + {3\over \sqrt 7} \sin \Bigl({\sqrt 7 \over 2}\ln 
{1\over z}\Bigr)\biggr]. 
\label{eq:defr}
\ee

However, the fact that in the charged current case the boson-gluon fusion 
process leads to a charm quark plus a light quark, rather than the charm 
quark-antiquark pair of the neutral current, leads to a technical complication.
For the neutral current the LO boson-gluon fusion coefficient function is 
infrared finite, and since it corresponds to the production of two massive 
quarks, vanishes smoothly at the kinematic threshold of $\hat W^2=4m_c^2$,
where $\hat W^2=Q^2(1/z-1)$. In contrast the charged-current boson-gluon
fusion has a collinear divergence due to the final state light quark which
must be regularized using dimensional regularization, and a subtraction made 
according to the rules of collinear factorization. The remaining finite 
coefficient function is no longer a real cross-section, and although it 
vanishes for $\hat W^2$ below the threshold of $\hat W^2=m_c^2$, 
it does not tend to zero
at the threshold, and in fact is logarithmically divergent as 
$\hat W^2 \to m_c^2$. The non-vanishing means that when taking the 
derivative of the right-hand side of eq(\ref{eq:n3}) with respect to $\ln Q^2$ 
one must also take 
account of the end-point of the derivative, i.e. we actually replace 
eq(\ref{eq:naive}) by
\be
\frac{d}{d \ln Q^2}
(C_{\s23 ,g}^{(1)FF}(\epsilon)\:\otimes\:\tilde g(Q^2))
= 
C_{\s23 ,\bar{\hs} c}^{(0)VF}(\epsilon) \:\otimes\: P^{(0)}_{qg}
\:\otimes\:\tilde g(Q^2),
\label{eq:real}
\ee 
The fact that the 
coefficient function is divergent at this point means the 
end-point contribution must be treated with particular care.   

Thus, in order to define $C_{\s23 ,\bar{\hs} c}^{(0)VF}(z,\epsilon)$ 
we separate out the part of the gluon coefficient function which diverges 
as $z \rightarrow x_0$ by writing
$$
C_{\s23 ,g}^{(1)FF}(z,\epsilon) = C_{\s23 ,g,reg.}^{(1)FF}(z,\epsilon) 
+ C_{\s23 ,g,dvgt.}^{(1)FF.}(z,\epsilon).
$$
This results in 
\bea
C_{2,g,reg.}^{(1)FF}(z,\epsilon) &=& \frac{1}{x_0} \left \{  
2P_{qg}^{(0)}(\zx0) [ L_{\lambda}(\zx0) -\ln x_0 ] \right \}\nonumber \\
 &+&  2[8-18(1-x_0) +12(1-x_0)^2]
\zx0(1-\zx0) + 2\left [ \frac{1-x_0}{1-z} -1 \right ] \nonumber \\
 &+& 2P_{qg}^{(0)}(\zx0) \ln \left [ \frac{x_0}{(1-z)z} \right ]
\; \left \{ +12(1-x_0)z (1-2z)L_{\lambda}(\zx0)  \right  \}, 
\label{eq:n5}
\eea
where
$L_{\lambda}(z) = \ln \left [ \frac{x_0(1-z)}{(1-x_0)z} \right ]$,
and
\be
C_{2,g,dvgt.}^{(1)FF}(z,\epsilon) = \frac{4}{x_0} P_{qg}^{(0)}(\zx0) 
\ln(1-\zx0). 
\label{eq:n6}
\ee
Also
\bea
C_{3,g,reg.}^{(1)FF}(z,\epsilon) &=&  
2P_{qg}^{(0)}(\zx0) [ -L_{\lambda}(\zx0) -\ln x_0 ] 
+4(1-x_0)\zx0(1-\zx0)  \nonumber \\
&+& 2(1-x_0)\zx0 L_{\lambda}(\zx0)\left [-2(1-\zx0)+2z \right ] \nonumber \\
&+& 2P_{qg}^{(0)}(\zx0) \ln \left [ \frac{x_0}{(1-z)z} \right ]
\label{eq:n7}
\eea
and
\be
C_{3,g,dvgt.}^{(1)FF}(z,\epsilon) = x_0 C_{2,g,dvgt.}^{(1)FF}(z,\epsilon) 
\label{eq:n8}
\ee

\medskip

It is easy to check that these coefficient functions approach the appropriate 
limits as $Q^2/m_c^2 \rightarrow \infty$. 
In this limit $\epsilon \rightarrow 0$, $x_0 \rightarrow 1$ and
\be
C_{2,g}^{(1)FF}(z,\epsilon) \rightarrow 2 \left \{ 8z(1-z)-1 + 
2P_{qg}^{(0)}(z) \ln \left [ \left (\frac{1-z}{z} \right )
\right ] +P_{qg}^{(0)}(z)\ln(1/\epsilon)\right \}
\label{eq:n9}
\ee
and
\be
C_{3,g}^{(1)FF}(z,\epsilon) \rightarrow - P_{qg}^{(0)}(z) 
\ln (1/\epsilon). 
\label{eq:n10}
\ee
Therefore both coefficient functions approach the massless form plus the 
appropriate collinear logarithms for the absorption into charm evolution.

\medskip

In the VFNS, $Q > m_c^2$, we need 
\bea
C_{\s23,\bar{\hs}c}^{(0)VF}(\epsilon) \otimes P_{qg}^{(0)} 
\:\otimes\:\tilde g(Q^2) &=&
\frac{d}{d\ln Q^2} \left [C_{\s23 ,g}^{(1)FF}(\epsilon)
\:\otimes\:\tilde g(Q^2)\right ] \nonumber \\
&=& \frac{d}{d\ln Q^2} \bigl(\left [ C_{\s23 ,g,reg.}^{(1)FF}(\epsilon)
+  C_{\s23 ,g,dvgt.}^{(1)FF}(\epsilon)\:\otimes\:\tilde g(Q^2) \right ]\bigr).
\label{eq:n11}
\eea
Let us consider each of these derivatives in turn. Strictly the derivative
is of the convolution so the terms generated by
differentiating the end point of the integration must be included.
We first consider the regular piece. 
$$
(a)\;\;\;\;\;\;\;\;\; 
\frac{d}{d\ln Q^2} \left [ C_{2,g,reg.}^{(1)FF}(\epsilon)
\:\otimes\:\tilde g(Q^2)
\right ].
$$
This results in
\bea
&\:&\frac{d}{d\ln Q^2} \left [\int_x^{x_0}dz\:  
C_{2,g,reg.}^{(1)FF}(z,\epsilon)\: \tilde g(x/z)  \right ] \nonumber \\
&=& \epsilon x_0^2 C_{2,g,reg.}^{(1)FF}(x_0,\epsilon) \tilde g(x/x_0) 
+ \epsilon x_0^2 \int_x^{x_0}\; dz \;  \frac{d}{dx_0} 
[C_{2,g,reg.}^{(1)FF}(z,\epsilon)] \;\tilde g(x/z) \nonumber \\
&=& -\epsilon x_0 \ln[x_0(1-x_0)] \tilde g(x/x_0) 
+ \epsilon x_0^2 \int_x^{x_0}\; dz \; \frac{d}{dx_0} 
[C_{2,g,reg.}^{(1)FF}(z,\epsilon)] \;\tilde g(x/z)  
\label{eq:n12}
\eea
and using eq(\ref{eq:n5}) we get
\bea
\frac{d}{dx_0} [C_{2,g,reg.}^{(1)FF}(z,\epsilon)] & = & 
 \left [\frac{2}{x_0^2(1-x_0)} \right ] P_{qg}^{(0)}(\zx0)
- \frac{2}{x_0^2(1-z)} \nonumber \\
&+& \frac{12z(1-2z)}{x_0^2} [ 1-L_\lambda(\zx0) ] 
  +  \frac{(4zx_0-6z^2-x_0^2)}{x_0^4}\ln[\frac{x_0}{(1-z)z^2}] \nonumber \\
 & + & \frac{4z}{x_0^2}[3z-2(1+3z)x_0+3(1+2z)x_0^2]
\label{eq:n13}
\eea
Now we also consider the divergent piece. 
$$
(b)\;\;\;\;\;\;\;\;\; 
\frac{d}{d\ln Q^2} \left [ C_{2,g,dvgt.}^{(1)FF}(\epsilon) 
\:\otimes\:\tilde g(Q^2)\right ].
$$
Differentiating this we obtain
\bea
&\:&\frac{d}{d\ln Q^2} \left [ \int_x^{x_0-\delta} dz\: 
C_{2,g,dvgt.}^{(1)FF}(z,\epsilon) \: 
\tilde g(x/z)  \right ] \nonumber \\
&=& \epsilon x_0^2 C_{2,g,dvgt.}^{(1)FF}(x_0-\delta,\epsilon) \tilde 
g(x/(x_0-\delta))
+ \epsilon x_0^2 \int_x^{x_0-\delta}\; dz \;  \frac{d}{dx_0} 
[C_{2,g,dvgt.}^{(1)FF}(z,\epsilon)] \;\tilde g(x/z), 
\label{eq:n14}
\eea
where for the moment we have moved the upper limit of integration an 
infinitesimal amount $\delta$ below $x_0$.
Writing 
$$
C_{2,g,dvgt.}^{(1)FF}(z,\epsilon) = \phi(z,\epsilon)\ln(1-\zx0),
$$
we get
\bea
&\:&\frac{d}{d\ln Q^2} \left [\int_x^{x_0-\delta}\:dz  
C_{2,g,dvgt.}^{(1)FF}(z,\epsilon) 
\tilde g(x/z)  \right ] \nonumber \\
&=& \epsilon x_0^2 \phi(x_0,\epsilon) \tilde g(x/(x_0-\delta)) 
\ln(1-\frac{x_0-\delta}{x_0})
 +  \epsilon x_0^2 \int_z^{x_0-\delta} \;dz\; 
\frac{d}{dx_0}[\phi(z,\epsilon)\ln(1-\zx0)] \tilde g(x/z). 
\label{eq:n15}
\eea
Now 
\be
\frac{d}{dx_0}[\phi(z,\epsilon)\ln(1-\zx0)] = \frac{z}{x_0^2(1-\zx0)}
\phi(z,\epsilon) + \frac{d\phi(z,\epsilon)}{dx_0}\ln(1-\zx0)
\label{eq:n16}
\ee
and since, in this case, $\phi(z,\epsilon) = \frac{4}{x_0}P_{qg}^{(0)}(\zx0)$
then
$$
\frac{d\phi(z,\epsilon)}{dx_0}\ln(1-\zx0) = \frac{2}{x_0^4}(4z x_0 -6z^2
-x_0^2)\ln(1-\zx0)
$$
and this contribution to the convolution can be added to the regular
contribution given by eq(\ref{eq:n13}).
Now the first term on the rhs of eq(\ref{eq:n16}) inserted into 
eq(\ref{eq:n15}) gives
\bea
&\:&\epsilon\int_x^{x_0-\delta}\; \frac{dz}{1-\zx0} \; z \phi(z,\epsilon) 
\tilde g(x/z) \nonumber \\ 
&=&\epsilon x_0 \phi(x_0,\epsilon) \tilde g(x/x_0) \int_x^{x_0-\delta}
\frac{dz}{1-\zx0} + \epsilon \int_x^{x_0-\delta}\;\frac{dz}{1-\zx0}\;
[z\phi(z,\epsilon)\tilde g(x/z) - x_0 \phi(x_0,\epsilon) \tilde g(x/x_0)] 
\nonumber \\
&=& - \epsilon x_0^2 \phi(x_0,\epsilon)\tilde g(x/x_0)\ln(1-
\frac{x_0-\delta}{x_0})
+\epsilon x_0^2 \phi(x_0,\epsilon) \tilde g(\xx0) \ln(1-\xx0) \nonumber \\
& + &  \epsilon \int_x^{x_0-\delta}\;\frac{dz}{1-\zx0}
[z\phi(z,\epsilon)\tilde g(x/z) - x_0 \phi(x_0,\epsilon) 
\tilde g(x/x_0)].
\label{eq:n17}
\eea
The first term in eq(\ref{eq:n17}) cancels the first term in
eq(\ref{eq:n15}) (up to ${\cal O}(\delta)$), 
and hence all divergences cancel as $\delta \to 0$. 
Removing these two terms and now safely setting $\delta=0$ the second term 
can be added to the first term of
eq(\ref{eq:n12}) as the net `local' contribution to the convolution.

So (a) and (b) together give the following contributions :-
\bea
{\rm `local' term} &:&  \epsilon x_0 [2 \ln(1-\xx0) - \ln(x_0(1-x_0))]
\tilde g(\xx0), \\
{\rm `+' term} &:& \frac{2\epsilon}{x_0}\int_x^{x_0} \; \frac{dz}{1-\zx0}
\;[2zP_{qg}^{(0)}(\zx0) \tilde g(x/z) - x_0 \tilde g(\zx0)], \\
{\rm `regular' term} &:& \epsilon x_0^2 \int_x^{x_0} dz
\left \{ \frac{d}{dx_0}
C_{2,g,reg.}^{(1)}(z,\epsilon)\left |_{\rm (as\;given\;in\;eq(\ref{eq:n13}))}
\right . \right . \nonumber \\
&+&\left . \frac{2}{x_0^4}(4z x_0-6z^2-x_0^2)\ln(1-\zx0) \right \} 
\tilde g(x/z)\nonumber \\
&\equiv& \int_x^{x_0} dz
\biggl(\frac{d C_{2,g}^{(1)}(z,\epsilon)}{d \ln Q^2}\biggr)_{reg} 
\tilde g(x/z).
\label{eq:n20} 
\eea
As $Q^2/m_c^2 \rightarrow \infty$, 
the only surviving term comes from the regular 
piece which $\rightarrow 2P_{qg}^{(0)}(z)$, and hence clearly using 
eq(\ref{eq:naive}),  
$C_{2 ,\bar{\hs} c}^{(0)VF}(z,\epsilon) \to 2z\delta(1-z)$ in this limit.

In general we can use the three contributions to $\frac{d}{d\ln Q^2}
C_{2,g}^{(1)FF}(\epsilon) \:\otimes\: \tilde g(Q^2)$ to derive three
contributions to the charm quark coefficient function 
$C_{\bar{\hs}c}^{(0)VF}(z,\epsilon)$, convoluted with the charm density.  
The part coming from the `regular' term eq(27) contributes in the 
normal manner as in eq(\ref{eq:conv}). 
The part coming from the local term is even simpler, becoming  
\bea
C_{\s23,\bar{\hs}c}^{(0)\;VF\;{\rm loc}}(\epsilon)\otimes c(Q^2)
&=& - f_{\rm loc}(x,x_0) 
\biggl({x\over x_0}\biggr)^2 \: \frac{d c(x/x_0,Q^2)}
{d (x/x_0)} \nonumber \\
&+& 2 f_{\rm loc}(x,x_0)
\:{x\over x_0} \: c(x/x_0,Q^2) \nonumber \\
&-& {2 \over x_0}\int_{x}^{x_0} dz  f_{\rm loc}(x,x_0)
r(\zx0) \: {x\over z}
\: c(x/z,Q^2),
\label{eq:convloc} 
\eea
where
\be
 f_{\rm loc}(x,x_0)=\epsilon x_0 [2 \ln(1-\xx0) - \ln(x_0(1-x_0))].
\label{eq:locdef}
\ee
The part coming from the `+' term is the most complicated. For the first two 
terms in the expression of the form eq(\ref{eq:conv}) it is relatively
straightforward, i.e. we obtain
\bea
&-&\frac{2\epsilon}{x_0}\int_x^{x_0} \; \frac{dz}{1-\zx0}
\;\biggl[2zP_{qg}^{(0)}(\zx0) \biggl(\frac{x}{z}\biggr)^2 
\frac{d c(x/z)}{d(x/z)}
 - x_0 \biggl(\frac{x}{x_0}\biggr)^2 
\frac{d c(x/x_0)}{d(x/x_0)}\biggr] \nonumber \\
&+& 2\frac{2\epsilon}{x_0}\int_x^{x_0} \; \frac{dz}{1-\zx0}
\;[2zP_{qg}^{(0)}(\zx0) \tilde c(x/z) - x_0 \tilde c(x/x_0)].
\label{eq:convplus}
\eea
In the final term we obtain a double convolution of the form
\be
\frac{2\epsilon}{x_0}\int_x^{x_0} \; \frac{dz}{1-\zx0}\biggl[
\;2zP_{qg}^{(0)}(\zx0) \int^1_{x/z} dz'r(z')\tilde c(x/zz')
 - x_0 \int^1_{x/x_0}dz'r(z')\tilde c(x/x_0z')\biggr].
\label{eq:doubconv}
\ee
In principle this can be calculated, but it is convenient to 
make a change of variables and use $y,z$ rather than $z,z'$, where $y=zz'$.
Doing this the first term in eq(\ref{eq:doubconv}) becomes
\be
\frac{2\epsilon}{x_0}\int_x^{x_0} \;\tilde c(x/y)  \int^{x_0}_{y} 
\frac{dz}{z(1-\zx0)}\;2zP_{qg}^{(0)}(\zx0)r(\frac{y}{z}).
\label{eq:dc1}
\ee
Changing variable in the second term gives
\be
-\frac{2\epsilon}{x_0}\int_x^{x_0} \;\tilde c(x/y)  \int^{x_0}_{x} 
\frac{dz}{x_0(1-\zx0)}\;x_0 r(\frac{y}{x_0}).
\label{eq:dc2}
\ee
The second integral is conveniently cut into two at $z=y$, producing 
\be
-\frac{2\epsilon}{x_0}\int_x^{x_0} \;\tilde c(x/y)
\biggl(\int^{x_0}_{y} \frac{dz}{x_0(1-\zx0)}\;x_0 r(\frac{y}{x_0})+
\int^{y}_{x} \frac{dz}{x_0(1-\zx0)}\;x_0 r(\frac{y}{x_0})\biggr).
\label{eq:dcsplit}
\ee
Altogether eq(\ref{eq:dc1}) and first part of eq({\ref{eq:dcsplit}) gives a 
contribution of the form
\be
\frac{2\epsilon}{x_0}\int_x^{x_0} \;\tilde c(x/y) \int^{x_0}_{y} 
\frac{dz}{(1-\zx0)}[2P_{qg}^{(0)}(\zx0)r(\frac{y}{z})- r(\frac{y}{x_0})],
\label{eq:dcplus}
\ee 
while the second part of eq({\ref{eq:dcsplit}) gives a 
contribution of the form
\be
\frac{2\epsilon}{x_0}\int_x^{x_0} \; dy\:\tilde c(x/y)r(\frac{y}{x_0})
\ln\biggl(\frac{1-y/x_0}{1-x/x_0}\biggr),
\label{eq:loca}
\ee
where the second integral over $z$ has been performed explicitly. 
Eq(\ref{eq:dcplus}) defines the '+' part while eq(\ref{eq:loca}) effectively
joins the local part eq(\ref{eq:convloc}). 

Thus, we have all the ingredients to define 
$C_{\bar{\hs}c}^{(0)VF}(\epsilon) \: \otimes \: \tilde c(Q^2)$. In order to 
obtain the complete LO expression for the generation of $\bar{\hs}$ quarks 
in the 
VFNS we then have to add all these above ingredients to the LO FFNS 
expression frozen at $Q^2=m_c^2$ as explained in \cite{TR}. Therefore
\be
F^{LO,\bar{\hs}}_{2}(x,Q^2) =  \left ( \frac{\alpha_S(m_c^2)}{2\pi} \right ) 
C_{\s23 ,g}^{(1)FF}(1) \:\otimes\: \tilde g(m_c^2)
+ C_{\bar{\hs}c}^{(0)VF}(\epsilon) \: \otimes \: \tilde c(Q^2),
\label{eq:LOdef}
\ee
for $Q^2>m_c^2$, where
\bea
C_{\bar{\hs}c}^{(0)VF}(\epsilon) \: \otimes \: \tilde c(Q^2) &=&
- \int_{x}^{x_0} dz \, \biggl(\frac{d C_{\s23 ,g}^{(1)FF}(z,\epsilon)} 
{d \ln Q^2}\biggr)_{reg}
\biggl({x\over z}\biggr)^2 \: \frac{d c(x/z,Q^2)}
{d (x/z)} \nonumber \\
&+& 2\int_{x}^{x_0} dx \, \biggl(\frac{d C_{\s23 ,g}^{(1)FF}(z,\epsilon)} 
{d \ln Q^2}\biggr)_{reg}
\:{x\over z} \: c(x/z,Q^2) \nonumber \\
&-& 2\int_{x}^{x_0} dz \, \biggl(\frac{d C_{\s23 ,g}^{(1)FF}(z,\epsilon)} 
{d \ln Q^2}\biggr)_{reg}
\int_{{x/z}}^1 d z'\, r(z') \: {x\over zz'}
\: c(x/zz',Q^2)\nonumber \\
&-& f_{\rm loc}(x,x_0) 
\biggl({x\over x_0}\biggr)^2 \: \frac{d c(x/x_0,Q^2)}
{d (x/x_0)} \nonumber \\
&+& 2 f_{\rm loc}(x,x_0)
\:{x\over x_0} \: c(x/x_0,Q^2) \nonumber \\
&-& {2 \over x_0}\int_{x}^{x_0} dz  f_{\rm loc}(x,x_0)
r(\zx0) \: {x\over z}
\: c(x/z,Q^2) \nonumber \\
&-&2\frac{2\epsilon}{x_0}\int_x^{x_0} \; dy\:\tilde c(x/y)r(\frac{y}{x_0})
\ln\biggl(\frac{1-y/x_0}{1-x/x_0}\biggr) \nonumber \\
&-&\frac{2\epsilon}{x_0}\int_x^{x_0} \; \frac{dz}{1-\zx0}
\;\biggl[2zP_{qg}^{(0)}(\zx0) \biggl(\frac{x}{z}\biggr)^2 
\frac{d c(x/z)}{d(x/z)}
 - x_0 \biggl(\frac{x}{x_0}\biggr)^2 
\frac{d c(x/x_0)}{d(x/x_0)}\biggr] \nonumber \\
&+& 2\frac{2\epsilon}{x_0}\int_x^{x_0} \; \frac{dz}{1-\zx0}
\;[2zP_{qg}^{(0)}(\zx0) \tilde c(x/z) - x_0 \tilde c(x/x_0)]\nonumber \\
&-& 2\frac{2\epsilon}{x_0}\int_x^{x_0} \;\tilde c(x/y) \int^{x_0}_{y} 
\frac{dz}{(1-\zx0)}[2P_{qg}^{(0)}(\zx0)r(\frac{y}{z})- r(\frac{y}{x_0})].
\label{eq:complete}
\eea
This expression then guarantees continuity of both 
the structure function and its derivative in $\ln Q^2$ as we switch 
from 3 to 4 flavours at $Q^2=m_c^2$.

\bigskip

Having completed the exercise for $F_2$ we can now do exactly the 
same thing for the phenomenologically interesting case of $F_3$. 
Once again we can first consider the contribution coming from the 
regular part of the FFNS coefficient function  
$$
(c)\;\;\;\;\;\;\;\;\; 
\frac{d}{d\ln Q^2} \left [ C_{3,g,reg.}^{(1)FF}(\epsilon) \:\otimes\: \tilde 
g(Q^2) \right ].
$$
Differentiating the contribution to the structure function due to 
this we obtain
\bea
&\:&\frac{d}{d\ln Q^2} \left [\int_x^{x_0}dz \:  
C_{3,g,reg.}^{(1)FF}(z,\epsilon) \: \tilde g(x/z)  \right ] \nonumber \\
&=& \epsilon x_0^2 C_{3,g,reg.}^{(1)FF}(x_0,\epsilon) \tilde g(x/x_0) 
+ \epsilon x_0^2 \int_x^{x_0}\; dz \;  \frac{d}{dx_0} 
[C_{3,g,reg.}^{(1)FF}(z,\epsilon)] \;\tilde g(x/z) \nonumber \\
&=& -\epsilon x_0^2 \ln[x_0(1-x_0)] \tilde g(x/x_0) 
+ \epsilon x_0^2 \int_x^{x_0}\; dz \; \frac{d}{dx_0} 
[C_{3,g,reg.}^{(1)FF}(z,\epsilon)] \;\tilde g(x/z), 
\label{eq:n21}
\eea
and using eq(\ref{eq:n7}) we get
\bea
\frac{d}{dx_0} [C_{3,g,reg.}^{(1)FF}(z,\epsilon)] & = & 
 \left [\frac{-2}{x_0(1-x_0)} \right ] P_{qg}^{(0)}(\zx0) \nonumber \\
&+& \frac{2z}{x_0^3}(x_0-2z) \ln \left[ \frac{x_0}{(1-x_0)z^2} \right] 
  +  \frac{4z}{x_0^3}(3z-2x_0). 
\label{eq:n22}
\eea
We can then also consider the divergent part of the structure function
$$
(d)\;\;\;\;\;\;\;\;\; 
\frac{d}{d\ln Q^2} \left [ C_{3,g,dvgt.}^{(1)FF}(\epsilon) 
\:\otimes\: \tilde g(Q^2)\right ].
$$
The only difference compared to (b) above is an extra factor of $x_0$ so that
$$
\frac{d\phi(z,\epsilon)}{d\ln Q^2} = \frac{4z}{x_0^3}(x_0-2z)
$$
So (c) and (d) together give the following contributions :-
\bea
{\rm `local' term} &:&  \epsilon x_0^2 [2 \ln(1-\xx0) - \ln(x_0(1-x_0))]
\tilde g(\xx0) \\
{\rm `+' term} &:& 2\epsilon \int_x^{x_0} \; \frac{dz}{1-\zx0}
\;[2zP_{qg}^{(0)}(\zx0) \tilde g(x/z) - x_0 \tilde g(\zx0)] \\
{\rm `regular' term} &:& \epsilon x_0^2 \int_x^{x_0}
\left \{ \frac{d}{dx_0}
C_{3,g,reg.}^{(1)}(z,\epsilon)\left |_{\rm (as\;given\;in\;eq(\ref{eq:n22}))}
\right . \right . \nonumber \\
&+&\left . \frac{4z}{x_0^3}(x_0-2z)\ln(1-\zx0) \right \} \tilde 
g(x/z) \nonumber \\
&\equiv& \int_x^{x_0} dz
\biggl(\frac{d C_{3,g}^{(1)}(z,\epsilon)}{d \ln Q^2}\biggr)_{reg} 
\tilde g(x/z).
\label{eq:n23} 
\eea
As $Q^2/m_c^2 \rightarrow \infty$, 
the only surviving term comes from the regular 
piece which $\rightarrow -2P_{qg}^{(0)}(z)$, and hence clearly using 
eq(\ref{eq:naive}),  
$C_{3 ,\bar{\hs} c}^{(0)VF}(\epsilon) \to -2z\delta(1-z)$ in this limit.
As for $F_2$ we can use the above three contributions to construct the 
necessary $C_{3,\bar{\hs}c}^{(0)VF}(\epsilon) \: \otimes \: \tilde c(Q^2)$
for the LO VFNS expression,
i.e. we obtain the equivalent of eq(\ref{eq:complete})
with the `regular' part eq(27) replaced by eq(43), and the `local' and 
`+' contributions being
identical to those for $F_2$ up to a factor of $x_0$.  

In practice, although it is convenient to talk about the production of $c$
quarks and or $\bar{\hs}$ quarks they are often produced together, and 
in order to define a physically relevant inclusive quantity we have to add the 
contributions we have considered for producing $\bar{\hs}$ quarks, i.e. 
eq(\ref{eq:n3}) in the FFNS and eq(\ref{eq:LOdef}) in the VFNS and their
analogues for
$F_3$, to the expressions for charm production eq(\ref{eq:n1}). Using these 
we have the contributions to the structure functions due to the production
and/or conversion of heavy flavours. The relevant curves are shown in   
figs. 1 and 2 for $F_2(x,Q^2)$ and $F_3(x,Q^2)$ respectively, 
where one can indeed see the continuity of both the structure 
functions and their derivatives, and the fact that they reduce to the correct 
limits at high and low $Q^2$.\footnote{We use the preliminary set of partons 
described in \cite{MRSTNNLO}.} (The small constant difference between the 
ZM-VFNS and the VFNS results at high $Q^2$ is due to the $m_c^2$-dependent 
first term on the rhs of eq(\ref{eq:LOdef}), which as we argued in 
\cite{TR} it is correct to include.) Note that at LO we have to use  
parton densities evolved according to only the LO splitting 
functions. 

\bigskip

\noindent{\large \bf 3 {\hskip 0.5cm} The VFNS at NLO.}

\medskip

We now consider the full range of NLO corrections to the charged-current
structure functions. 
There is in principle a next-to-leading order correction to the 
production of charm quarks from $\hs$ quarks, i.e.  
\be 
\left ( \frac{\alpha_S}{4\pi} \right )
\int_x^{x_0} \;dz \;C^{(1)}_{\s23 ,c\hs}(z,\epsilon)\; \bar{\hs}(x/z) 
\label{eq:n2}
\ee
where the $C_{\s23 ,c\hs}^{(1)}(z)$ are presented in \cite{gkr}. These 
coefficient functions have no large logs in $Q^2/m_c^2$ and simply reduce
to the correct massless expressions as $\epsilon \to 0$,
and are the same in VFNS as in FFNS. However, the
contribution from eq(\ref{eq:n2}) is essentially negligible at all $Q^2$ 
and $x$. Hence, we use massless coefficient function for this process for
simplicity. 

There are then also other contributions at NLO. These are due to the 
coefficient functions $C^{(2)FF}_{\s23 ,g}(z,\epsilon)$, 
$C^{(2)FF,PS}_{\s23 ,\bar{\hs}q}(z,\epsilon)$ (where $PS$ stands for pure 
singlet), $C^{(1)VF}_{\s23 ,\bar{\hs}c}(z,\epsilon)$ and 
$C^{(1)VF}_{\s23 ,g}(z,\epsilon)$. We will first consider the last
of these, since this is the easiest to deal with.    

The explicit form of eq({\ref{eq:partons}) for $\mu^2=Q^2$ at
${\cal O}(\alpha_s)$ is
\bea
c(z,Q^2) &=& \frac{\alpha_s}{2\pi}\;
\ln \biggl (\frac{Q^2}{m_c^2} \biggr )
\;P_{qg}^0 \:\otimes\: g_{n_f=3} \nonumber \\ 
g_{n_f=4}(z,Q^2) &=& g_{n_f=3}(z,Q^2) \:-\:\frac{\alpha_s}{6\pi}
\;\ln \biggl (\frac{Q^2}{m_c^2} \biggr )\; g_{n_f=3}. 
\label{eq:partdefcharm}
\eea
Inserting the expressions for the matrix element 
$A^{cg}(z,\mu^2/m_c^2)$ 
into eq(\ref{eq:corrcoeff}) gives the simple relation
\be
C_{\s23, g}^{(1)\;FF}(z,\epsilon) = 
C_{\s23, g}^{(1)\;VF}(z,\epsilon) 
\:+\: C_{\s23, \bar{\hs}c}^{(0)\;VF}(\epsilon) \:\otimes\: P_{qg}^0\;
\ln \biggl (\frac{1}{\epsilon} \biggr )
\label{eq:cfgconnect}
\ee
connecting the ${\cal O}(\alpha_S)$ gluonic CF's in the FFNS and VFNS. 
Futhermore 
eq(\ref{eq:real}) allows
the gluonic CF in the VFNS to be written as
\be
C_{\s23, g}^{(1)\;VF}(\epsilon)\:\otimes\: \tilde g(Q^2) = 
C_{\s23, g}^{(1)\;FF}(\epsilon) \:\otimes\: \tilde g(Q^2)\:-\:
\frac{d}{d \ln Q^2} (C_{\s23, g}^{(1)\;FF}(\epsilon)\:\otimes\: 
\tilde g(Q^2))\; 
\ln \biggl ( \frac{1}{\epsilon} \biggr ).
\label{eq:defnlogcf}
\ee   
Hence, it is a straightforward procedure to take the results of 
the previous section regarding the correct treatment of 
$\frac{d}{d \ln Q^2} (C_{\s23, g}^{(1)\;FF}(\epsilon)\:\otimes\: 
\tilde g(Q^2))$,
including the contributions from the endpoint of the integral in the 
convolution, to completely define 
$C_{\s23, g}^{(1)\;VF}(z,\epsilon)$. As with 
$C_{\s23, \bar{\hs}c}^{(0)\;VF}(z,\epsilon)$ there is a `regular', `local'
and `+' contribution. Using the asymptotic limits for the FFNS coefficient 
functions in eq(\ref{eq:n9}) and eq(\ref{eq:n10}), along with the limits on 
their $\ln Q^2$ derivatives, presented in the previous section,  
we see that as $Q^2/m_c^2 \rightarrow \infty$,
$C_{\s23, g}^{(1)\;VF}(z,\epsilon)$ do indeed tend to the 
correct asymptotic $\msb$ limit. 
   
In principle $C^{(2)FF}_{\s23 ,g}(z,\epsilon)$ and  
$C^{(2)FF,PS}_{\s23 ,\bar{\hs}q}(z,\epsilon)$ contribute at NLO. 
This is both in the FFNS expressions for $Q^2<m_c^2$, and in the VFNS
where the values frozen at $Q^2=m_c^2$ are used to ensure continuity of 
the structure function. Unfortunately, unlike the neutral current case
\cite{NLOnc}, 
neither of the contributions has been calculated yet, and as such we have 
no option but to leave them out completely. However, 
$C^{(2)FF}_{\s23 ,g}(z,\epsilon)$ also has a role to play in 
the definition of the NLO VFNS coefficient function 
$C^{(1)VF}_{\s23 ,\bar{\hs}c}(z,\epsilon)$, and it is not possible to 
simply claim ignorance and set this to zero since this would destroy the 
continuity of $\frac{d F_{\s23}(x, Q^2)}{d\ln Q^2}$ at NLO. 

To see this we must consider the equation defining 
$C^{(1)VF}_{\s23 ,\bar{\hs}c}(z,\epsilon)$. This is analogous to the case for
the neutral coupling discussed in section 4 of the former of \cite{TR}, and
we have the definition
\be
\frac{dC^{(2)FF}_{\s23,g}(z,\epsilon)}{d\ln Q^2} = 
C^{(1)VF}_{\s23,\bar{\hs}c}(\epsilon) \:\otimes\: \frac{d
A^{(1)}_{cg}(\epsilon)}{d \ln Q^2} 
+ C^{(0)VF}_{\s23,\bar{\hs}c}(\epsilon) \: \otimes \: \frac{d
A^{(2)}_{cg}(\epsilon)}{d \ln Q^2} 
+ \frac{1}{3\pi}\ln(1/\epsilon) C^{(0)VF}_{\s23,\bar{\hs}c}(\epsilon) \:
\otimes \: P^{(0)}_{qg},
\label{eq:NLOcoeff}                
\ee  
where the last term comes about from the difference in the definition of the
three and four flavour couplings. This expression would guarantee both the
continuity of the $\ln Q^2$-derivative of the structure function at NLO 
(in the gluon sector), and the correct asymptotic expression for
$C^{(1)VF}_{\s23,\bar{\hs}c}(z,\epsilon)$ - all terms containing a power of
$\ln(1/\epsilon)$ being guaranteed to cancel. However, since we do not know 
the NLO FFNS coefficient function, we cannot therefore fully use the above 
equation. Nevertheless, simply putting 
$C^{(1)VF}_{\s23,\bar{\hs}c}(z,\epsilon)$ 
equal to its asymptotic value, which for the moment we consider to be in
practice zero,  
is not consistent since this leads to the right-hand side of
eq(\ref{eq:NLOcoeff}) being equal to
\be
C^{(0)VF}_{\s23,\bar{\hs}c}(1)\: \otimes \: P^{(1)}_{qg}
\label{eq:mismatch}
\ee
at $Q^2=m_c^2$, where we have used the expression for $\frac{d
A^{(2)}_{cg}(\epsilon)}{d \ln Q^2}$ in eq(4.15) of the former of \cite{TR}, 
whereas the left-hand side is zero. Thus, there is a mismatch 
between the lack of evolution at NLO for $Q^2< m_c^2$,
and from the NLO contribution 
to the evolution from the NLO quark-gluon splitting function for $Q^2>m_c^2$
convoluted with the zeroth order coefficient function. This mismatch 
may be large, particularly at small $x$. 
In order to avoid this we have to invoke some ansatz for
$C^{(1)VF}_{\s23,\bar{\hs}c}(\epsilon)$ so that the above contribution is
cancelled. Using the fact that $\frac{d A^{(1)}_{cg}(z,\epsilon)}{d \ln Q^2}
= P^{(0)}_{qg}(z)$, this results in the requirement
\be
C^{(1)VF}_{\s23,\bar{\hs}c}(1)\:\otimes\:P^{(0)}_{qg}
+C^{(0)VF}_{\s23,\bar{\hs}c}(1) \:\otimes\: P^{(1)}_{qg} =0.
\label{eq:ansatz}
\ee
The minimal way in which to satisfy this, and to ensure that
$C^{(1)VF}_{\s23,\bar{\hs}c}(\epsilon) \to 0$ as $\epsilon \to 0$, is to    
demand that
\be
C^{(1)VF}_{\s23,\bar{\hs}c}(\epsilon)\:\otimes\: P^{(0)}_{qg} = -\epsilon
\: C^{(0)VF}_{\s23,\bar{\hs}c}(\epsilon)\:\otimes\: P^{(1)}_{qg}.
\label{eq:cnlodef}
\ee

In principle this is the definition we use for
$C^{(1)VF}_{\s23,\bar{\hs}}(z,\epsilon)$, but this would be extremely 
complicated to implement in practice. Since the coefficient function is always
convoluted with a parton distribution, and is based on the known 
$C^{(0)VF}_{\s23,\bar{\hs}c}(z,\epsilon)$ we find an appropriate modification 
of $C^{(0)VF}_{\s23,\bar{\hs}}(z,\epsilon)$ necessary to account for the effect
of the NLO coefficient function. We find that assuming that the parton
distribution takes roughly the form $(1-x)^8x^{-0.3}$ then we can model 
the action of this NLO coefficient function by replacing all terms of the form 
\be
\frac{d C^{(1)FF}_{i,g}(z,\epsilon)}{d \ln Q^2} \:\otimes\: \tilde f(x/z, Q^2)
\label{eq:NLO1}
\ee
occurring in eq(\ref{eq:complete}) by
\be
\frac{d C^{(1)FF}_{i,g}(z,\epsilon)}{d \ln Q^2} \:\otimes\: \biggl(1-\epsilon
\frac{38\alpha_S(Q^2)}{4\pi}(\ln(4+(x/z)^{-0.25})-\ln(4)-2(x/z)\biggr)
\tilde f(x/z,Q^2),
\label{eq:NLOex}
\ee
with analogous modifications for the `local' and `$+$' contributions.
It can be checked explicitly that this does indeed 
represent the exact expression
eq({\ref{eq:cnlodef}) very accurately. The main effect is an opposite sign 
correction to the LO result which increases in magnitude as one goes to 
smaller $x$. Examining 
eq({\ref{eq:cnlodef}) one sees that $C^{(1)VF}_{\s23,\bar{\hs}c}$ depends on 
$P^{(1)}_{qg}/P^{(0)}_{qg}$ (where the division is really only illustrative
since convolutions are involved), and since $P^{(1)}_{qg}$ grows much more
quickly at small $x$ than $P^{(0)}_{qg}$ this effect is fully expected. 
Finally, we also add a contribution of $(1-\epsilon)
C^{(1)ZM-VF}_{i,\bar{\hs}c}$ in order to obtain the 
correct asymptotic limit, though the contribution due to this is tiny at small
$Q^2$.

This completes our definition of the VFNS for charged current scattering at 
NLO. Unfortunately a complete definition will have to await the calculation of
the unknown NLO FFNS coefficient functions, but we are confident that this will
lead to only small corrections, mainly for $Q^2\lapproxeq m_c^2$. 
In figs. 3 and 4 we plot the
contributions to $F_2(x,Q^2)$ and $F_3(x,Q^2)$ respectively due to the 
production and/or conversion of charm quarks at NLO.\footnote{Again using 
partons from \cite{MRSTNNLO}.} Once again one can see the continuity of the
structure functions and their derivatives, and the correct asymptotic 
behaviour\footnote{This time the high $Q^2$ limits of the VFNS and the ZM-VFNS
are identical since the constant difference would depend on the unknown NLO
FFNS coefficient functions} (we plot the FFNS result obtained from the 
LO coefficient functions since those at NLO are not known). This time we use 
partons evolved at NLO. For comparison we also plot the NLO structure 
functions with $C^{(1)VF}_{\s23,\bar{\hs}c}$ set equal to zero. 
One can see that at small $x$
this does indeed lead to a clear discontinuity in the derivative in the
structure function at the transition point $Q^2=m_c^2$, particularly as one 
goes to smaller $x$. This is, however, far more clear for $F_3(x,Q^2)$ where
we are calculating roughly the difference between the strange and charm quark 
distributions and the discrepancy in the evolution affecting one shows up 
much more obviously than for $F_2(x,Q^2)$, which is roughly the sum of the 
two quark distributions.        
   
One can extend the treatment to higher orders in principle following the 
general outline provided in the former of \cite{TR}. As mentioned in the 
introduction, at NNLO and beyond there is 
a complication in so much that particular flavours may be generated in the
final state due to the cutting of quark loops produced away from the
interaction vertex with the external gauge boson. This highlights the 
experimental ambiguity in defining heavy flavour structure functions and in
principle one needs define some kinematic cut on such quarks to decide whether
they are included or not. This issue is treated in \cite{smith} for the neutral
current case, though in practice the effect is extremely small. Since for
charged currents we do not even have a complete definition of the VFNS or 
FFNS at NLO this issue is not particularly pressing at the moment.   

\bigskip

\noindent{\large \bf 4 {\hskip 0.5cm} $\Delta xF_3(x,Q^2)$}  

\medskip
 
As we can see from figs. 1--4  our VFNS works well, ensuring smoothness and
the correct limits. We can repeat exactly the same procedure for the process
$W^- \to \bar c,\hs$, and this then allows the calculation of combined 
neutrino and antineutrino cross-sections, as measured by CCFR, and also the
currently interesting quantity $\Delta xF_3 =xF^{\nu N}_3-xF^{\bar\nu N}_3$
\cite{unki} where $N$ represents an isoscalar target. 
We present our results for 
$\Delta xF_3$ in fig. 5, for the range of $x$ relevant for the CCFR 
experiment ($x>0.01$). The curves are extremely similar to those for
$xF_3(x,Q^2)$ (with a factor of two), and would show the same type of kink 
at low $x$ if $C^{(1)VF}_{3,\bar{\hs}c,(\bar c \hs)}(z,\epsilon)$ were 
set equal to its asymptotic value or to zero. We also present the data on 
$\Delta xF_3$ measured by CCFR \cite{unki}, and note that our predictions lie
considerably beneath the measurements. Possible reasons for this are 
considered in \cite{kosty}. Since the data is at quite low $Q^2$ it is 
clear that the FFNS would lead to very similar predictions. By comparison 
with fig. 4 one can see that there would be a 
very slight improvement for the higher $Q^2$ points, but only due to missing
contributions correctly accounted for in the VFNS. Similarly the ZM-VFNS 
would actually compare to the data fairly well, but is simply incorrect
at such low $Q^2$.    

\bigskip

\noindent{\large \bf 5 {\hskip 0.5cm} Summary}

\medskip

In this paper we have explicitly constructed a VFNS for the production and
conversion of heavy flavours for the case of charged currents. 
We have demonstrated that the predicted structure function is very well 
described over a wide range of $x$ and $Q^2$ - having the correct asymptotic
limits for low $Q^2$ and for $Q^2 \to \infty$. We note that a VFNS is
particularly important in this case. For $Q^2\sim m_c^2$ there is no reason why
the ZM-VFNS should be a particularly good approximation to the correct 
structure function since it is missing essential information on the 
kinematics. Indeed, it is not that successful in the neutral current case,
often leading to a negative $F_2^c$ for low $Q^2$ and being much too high for
$F_L^c$ as seen in e.g. the first of \cite{TR}. 
It has, however, been argued, e.g. \cite{GRS}, that the FFNS is sufficient 
even up to $Q^2>>m_c^2$, and for the neutral current $F_2$  
it seems arguable that this is correct
(particularly if the renormalization/factorization scale is chosen
judiciously). However, it was demonstrated in \cite{vanneerven} that
particularly for the case of $F_3$, which is best measured in neutrino
scattering, this is no longer true, and at high $Q^2$ the FFNS expansion is
very slow to converge towards a resummed VFNS result and changes considerably 
from order to order. Hence, in this case the FFNS is
clearly unreliable at high $Q^2$ and a VFNS is needed.

Our particular scheme is built upon 
two basic ideas - incorporation of the correct kinematic behaviour into
each coefficient function by imposition of the continuity of $(d F(x,Q^2)/d \ln
Q^2)$ across the transition point $Q^2=m_c^2$, and a correct ordering of the
expansion in $\alpha_S$, so that a well-defined expansion scheme is used in
each limit and in between. However, these two ideas are linked by the complete
definition. In the case of charged currents the former no
longer appears to be such a direct benefit as for the neutral current case, 
because even the
lowest order boson-gluon fusion diagram needs a collinear subtraction due to
the final state light quark, and thus the finite part is not a true parton
cross-section. This means that unlike for the quark-antiquark production in the
neutral current case the coefficient function does not vanish at threshold, 
and is even divergent. This leads both to technical difficulty, with our
coefficient functions containing `$+$' distributions, and to there being a 
less direct link between the coefficient functions and the physics. As such,
superficially there seems to be no advantage compared to other VFNSs. 
Nevertheless, the
ordering still remains an advantage. Not only is it 
theoretically correct, combining renormalization/factorization scheme
independence up to higher orders with continuity of structure functions, but
it has a clear phenomenological benefit. This becomes particularly clear at 
NLO, where the ordering and the continuity of the $\ln Q^2$-derivative of the
structure function impose conditions on $C^{(1)VF}_{i,\bar{\hs}c}$ even in the
absence of the NLO FFNS coefficient functions, forcing smoothness by relating
this NLO coefficient function to the NLO evolution.  
Other schemes, e.g. \cite{acot,smith},
do not have the same type of definition of $C^{(1)VF}_{i,\bar{\hs}c}$,
i.e. do not relate it to $P^{(1)}_{qg}$
and would, we believe, have similar behaviour to our curves with this
coefficient function set equal to zero if the scale $\mu^2=Q^2$ were used. 
This unphysical behaviour would, however, be reduced if arguably more 
physical scales, such as $\mu^2=Q^2+m_c^2$ were used. 

Along with this paper we will make available new code for calculating
the heavy flavour contribution to charged currents.\footnote{The FORTRAN code
for this prescription can be obtained from 
{\tt http://durpdg.dur.ac.uk/HEPDATA/PDF}.} This contains various 
changes and corrections compared to the previous version. In particular 
we no longer use the coefficient function in \cite{witten} for calculation of
$F_2(x,Q^2)$ in the charged current case, since this seems to be incompatible
with those in \cite{gkr} and \cite{gott}, and we choose to believe these
since \cite{gkr} has been extensively cross-checked.\footnote{We would like
to thank Stefan Kretzer for details on this.} This change in 
coefficient functions leads to a significant
reduction in $F_2(x,Q^2)$ at low $Q^2$, though the difference disappear at
high $Q^2$. Thus, we now have a complete, explicit prescription for the
production of charged current structure functions including heavy flavour
effects which may be used along with LO or NLO $\overline{\rm MS}$ partons
distributions. We hope this will prove useful to the community.

\bigskip

\noindent{\large \bf {\hskip 0.5cm} Acknowledgements.}

\medskip

We would like to thank Stefan Kretzer, Fred Olness, Un-ki Yang and Randy
Scalise for useful conversations and communications, and Un-ki Yang for 
drawing to our attention some of the problems with our original code. 
RST would also like to thank the Royal Society for financial support.

\newpage

\begin{figure}[H]
\vspace{-1.0cm}
\begin{center}
\epsfig{figure=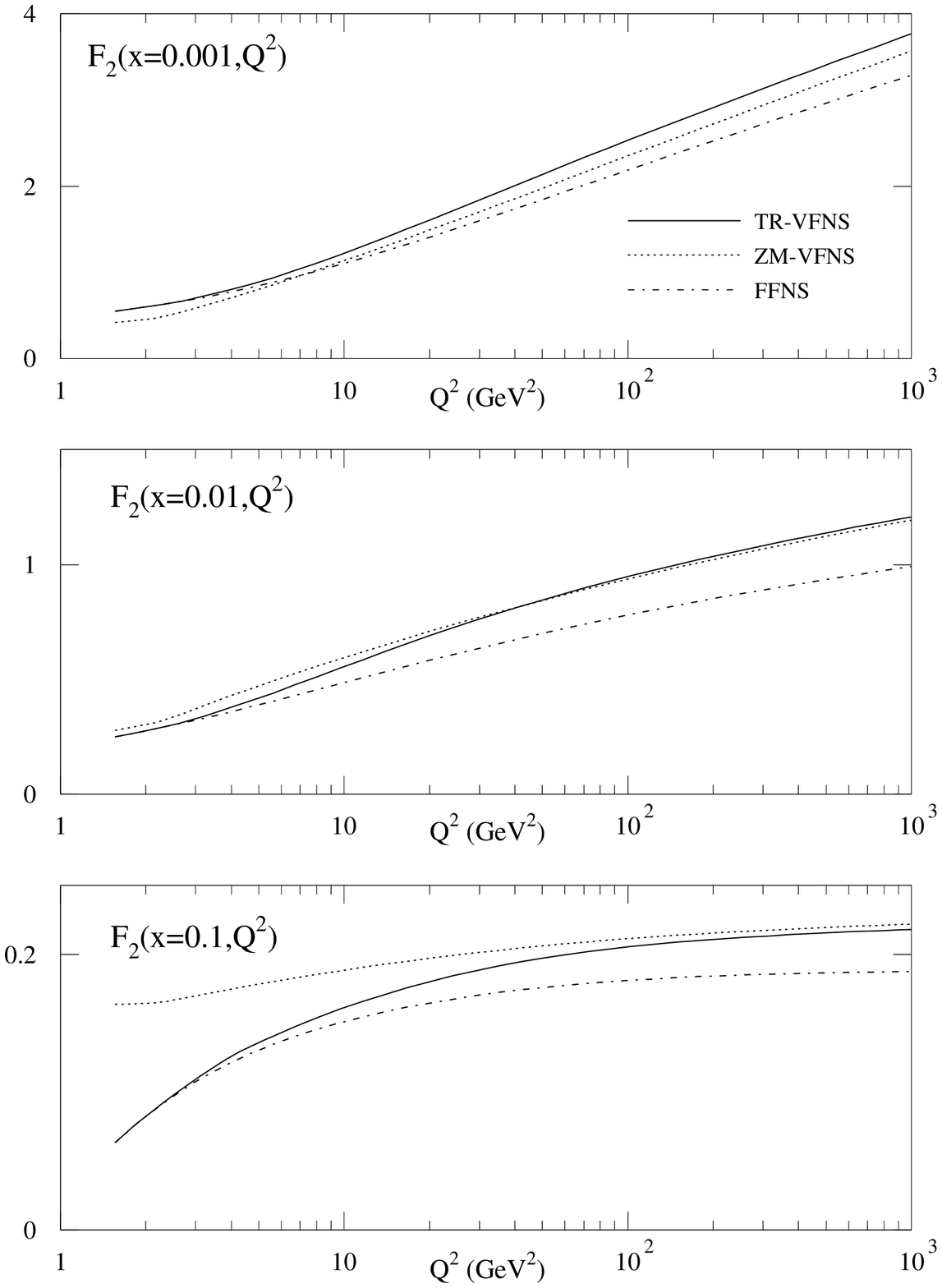,height=20cm}
\end{center}
\caption{Charm quark contribution to the
structure functions, $F_{2}(x,Q^2)$ for $x=0.1$,
$x=0.01$ and $x=0.001$ calculated using our LO prescription, our input parton 
distributions evolved at LO and renormalization scale $\mu^2=Q^2$. Also shown 
are the continuation of the LO FFNS expression and the ZM--VFNS expression 
both calculated using the same parton distributions and same choice of scale.}
\label{fig:Fig1}
\end{figure}

\begin{figure}[H]
\vspace{-1.0cm}
\begin{center}
\epsfig{figure=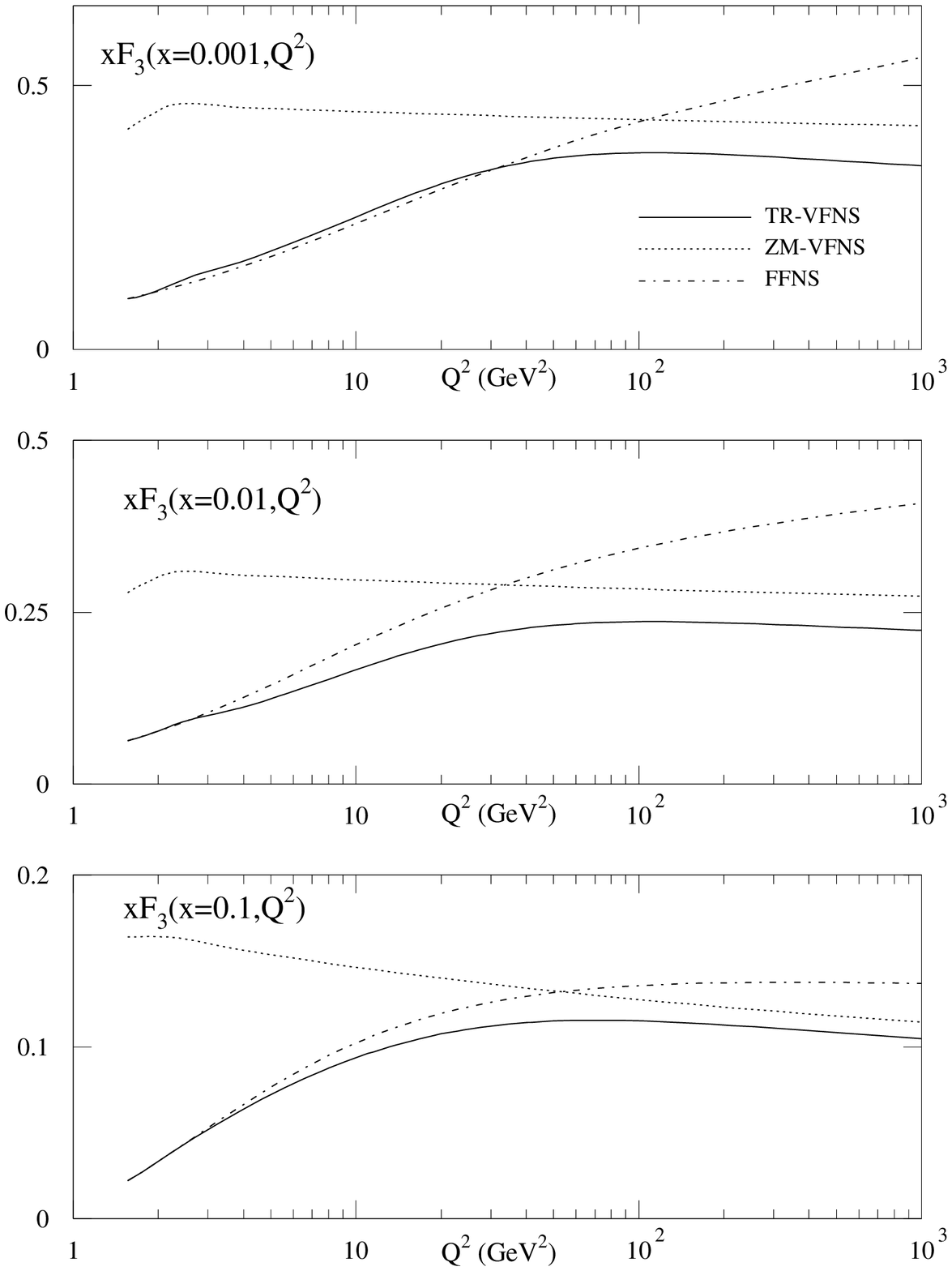,height=20cm}
\end{center}
\caption{Same as Fig. 1, but for $F_3(x,Q^2)$.}
\label{fig:Fig2}
\end{figure}

\begin{figure}[H]
\vspace{-1.0cm}
\begin{center}
\epsfig{figure=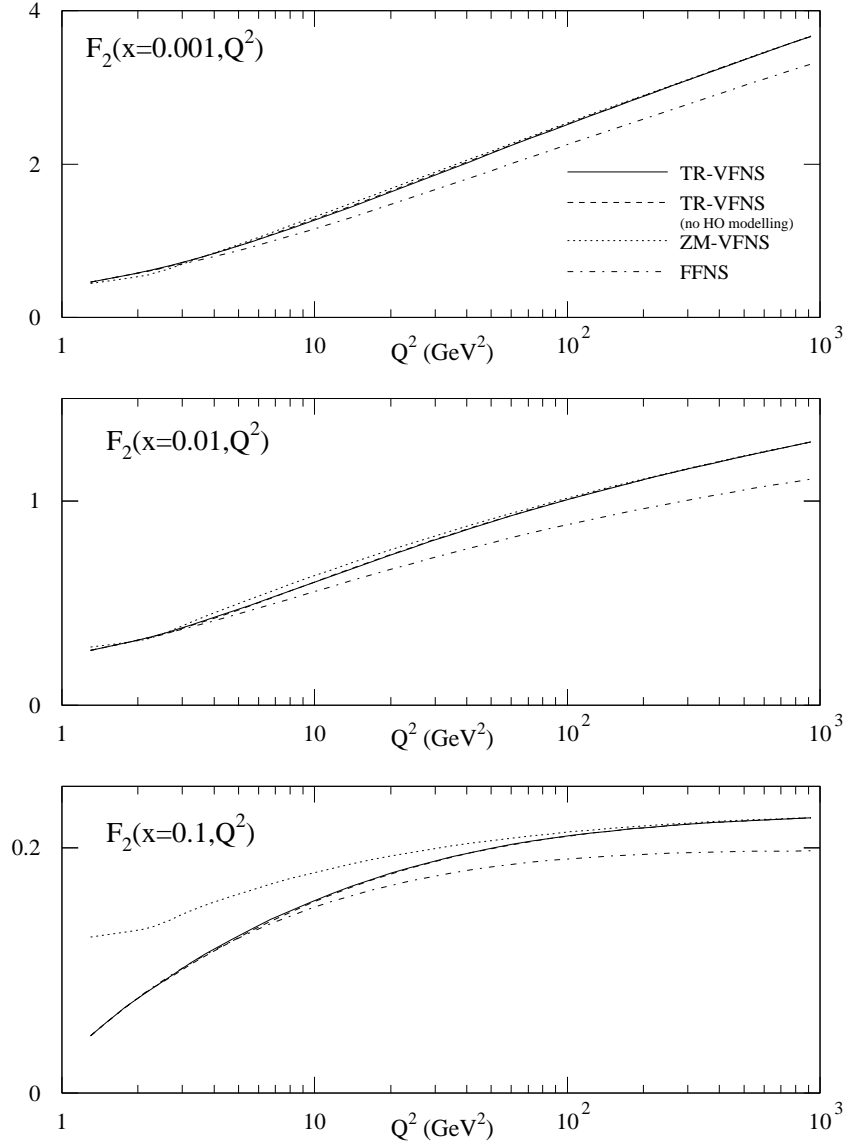,height=18cm}
\end{center}
\caption{Charm quark contribution to the
structure functions, $F_{2}(x,Q^2)$ for $x=0.1$,
$x=0.01$ and $x=0.001$ calculated using our NLO prescription, our input parton 
distributions evolved at NLO and renormalization scale $\mu^2=Q^2$. Also shown 
are the continuation of the FFNS expression with LO coefficient functions 
(those at NLO being unavailable) and the NLO ZM--VFNS expression 
both calculated using the same parton distributions and same choice of scale.
Also shown for comparison is the VFNS result when 
$C^{(1)VF}_{2,\bar{\hs}}$ is set equal to zero.} 
\label{fig:Fig3}
\end{figure}

\begin{figure}[H]
\vspace{-1.0cm}
\begin{center}
\epsfig{figure=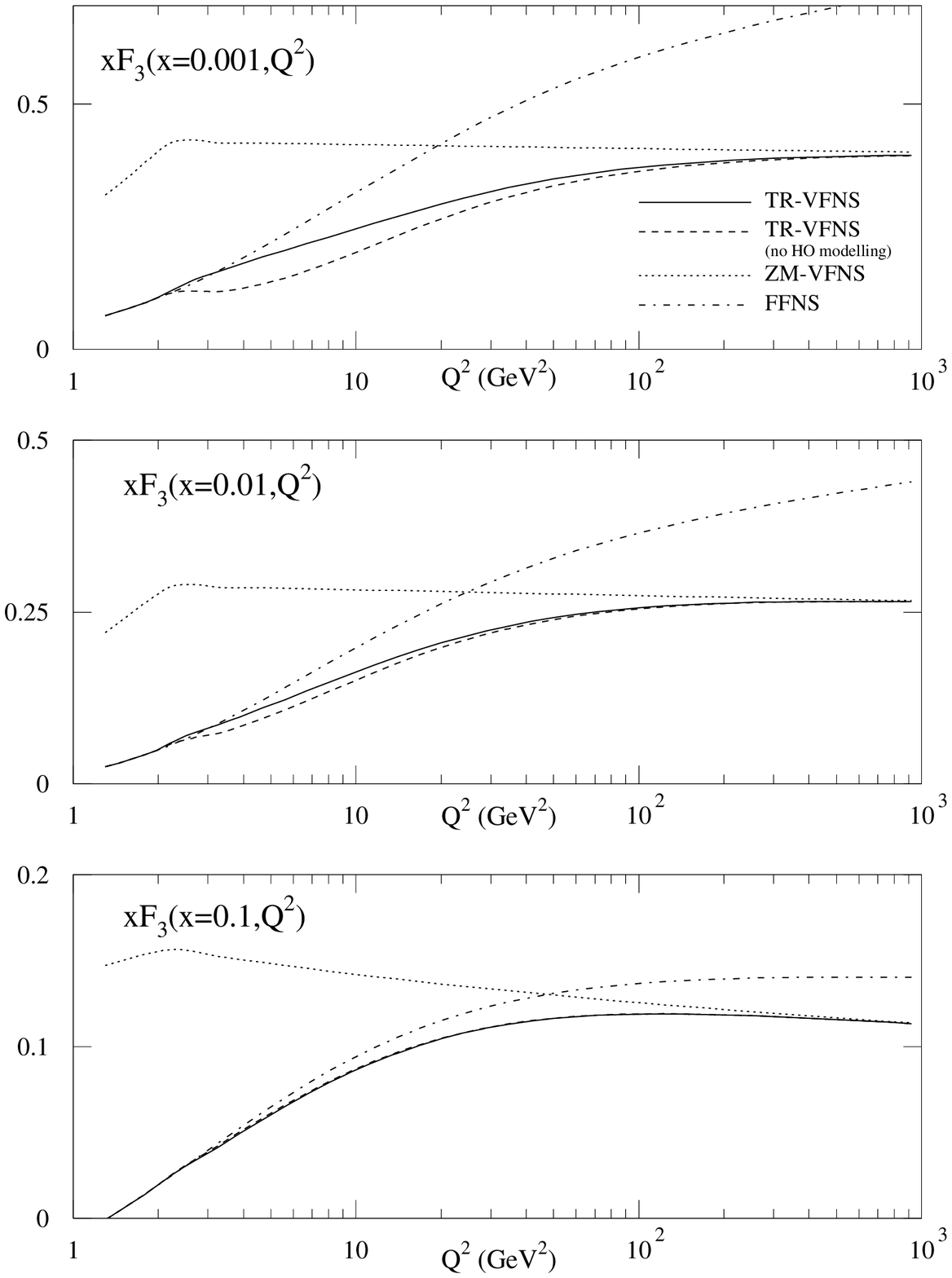,height=20cm}
\end{center}
\caption{Same as fig. 3, but for $F_3(x,Q^2)$.}
\label{fig:Fig4}
\end{figure}

\begin{figure}[H]
\vspace{-1.0cm}
\begin{center}
\epsfig{figure=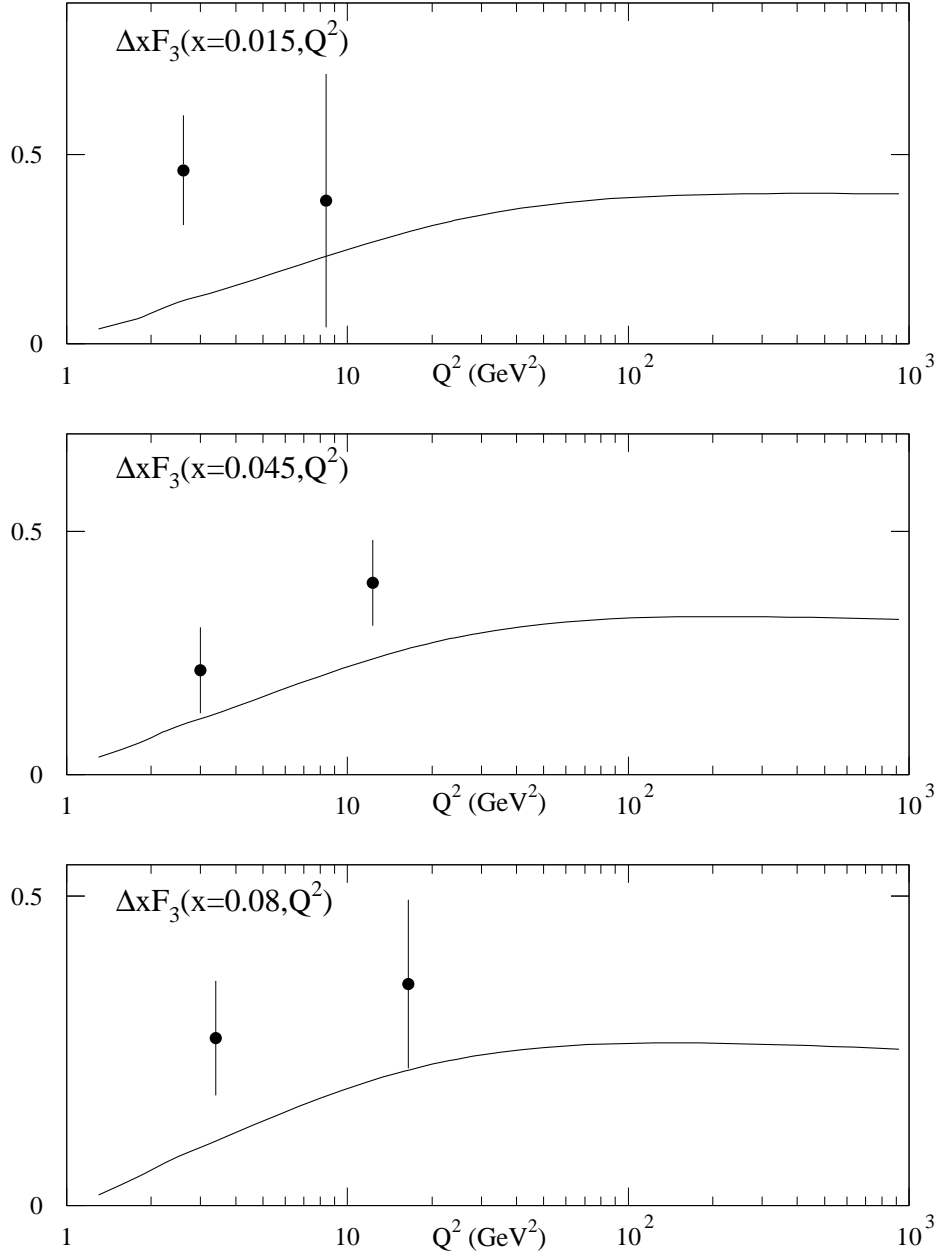,height=20cm}
\end{center}
\caption{The NLO prediction for $\Delta xF_3(x,Q^2)$ using our VFNS 
prescription, along with the data measured by CCFR \protect\cite{unki}. 
The prediction has been corrected for heavy target effects 
using \protect\cite{badkwie}.}
\label{fig:Fig5}
\end{figure}

\end{document}